\documentclass[intlimits,twoside,a4paper]{article}

\usepackage{amsmath,amssymb}
\usepackage{graphicx}
\usepackage{psfrag}
\usepackage[T2A]{fontenc}
\usepackage[cp1251]{inputenc}

\usepackage[eqsecnum]{cmpj2}

\issue{2016}{19}{3}{33701}
\doinumber{10.5488/CMP.19.33701}

\title[Magnetobreakdown oscillations of Nernst-Ettingshausen field in layered conductors]
{Magnetobreakdown oscillations of Nernst-Ettingshausen field in layered conductors}

\author[O.~Galbova]{O.~Galbova\thanks{E-mail: galbova@pmf.ukim.mk}}

\address{Faculty of Natural Sciences and
Mathematics, Institute of Physics, \\P.O. Box 162 1001 Skopje, Republic of Macedonia}

\authorcopyright{O.~Galbova, 2016}
\date{Received February 23, 2016, in final form April 25, 2016}

\begin{document}

\maketitle

\begin{abstract}

In the presented report, the Nernst-Ettingshausen effect in layered conductors is investigated. Considering a Fermi surface (FS) consisting of a slightly corrugated cylinder and two corrugated planes distributed periodically in the momentum space, the thermoelectric effects are considered under general assumptions for the value of a magnetic breakdown probability.  As a result of an external generalized force, the FS sheets in layered conductors with a multisheet FS appear to be so close that the charge carriers (as a result of magnetic breakdown) can move from one FS sheet to another. In addition, the distribution functions of the charge carriers and the magnetic breakdown oscillations of thermoelectrical field along the normal to the layer, under different values and orientations of the magnetic field, $B$, are calculated. It is shown that if the magnetic field is deflected from the $xz$-plane at an angle $\varphi$, the oscillation part of a thermoelectrical field along the normal to the layer under condition $\sin\varphi\tan\vartheta \gg 1$ is mainly determined with the Nernst-Ettingshausen effect.


\pacs 72.15.Gd, 74.70.Kn
\end{abstract}

\section{Introduction}

The electronic phenomena that occur in degenerate conductors, in the presence of strong magnetic fields, are highly dependent on the electron energy spectrum. The experimental study of these phenomena enables us to gain important information on the topology of the Fermi surface (FS) --- a basic/fundamental characteristic of the electron energy spectrum. Theoretical investigations of the metal magnetic permeability, under general assumptions for the type of energy spectrum of the conduction electrons (taken  a priori as the known one), were done by Lifshicz and Kosevich \cite{1}. It was shown that the period of oscillations of magnetization in a quantized magnetic field, $B$, as a function $1/B$ is proportional to the extreme FS section. The investigation of these oscillations, under different orientations of a magnetic field, allows for a complete determination of the form of the Fermi surface \cite{2}. The analogous information of the FS could be gained from the investigation of the Shubnikov de Haas magneto-resistance oscillations \cite{3,3a} in degenerate conductors \cite{4}. The investigation of galvanomagnetic phenomena in a strong magnetic field (classical case), where the circular frequency $\omega_\textrm{c}={eH}/(m^*c)$ is much bigger than the relaxation frequency, $1/\tau$, enables one to determine the FS topology structure \cite{5,6}. In the low-dimensional conductors, the oscillation effects are numerous.

Experimental observations (1998) performed in the Shegoleva (Chernogolovka) laboratory, at magnetic field strengths of up to 14~T, show the existence of resistance oscillations with a varying angle between the magnetic field, $\vec{H}$, and the normal to the layer, $\vec{n}$, in the layered organic conductor $\beta\text{-}(\text{BEDT}$-TTF)$_2\text{JBr}_2$ \cite{7,8}. Then, the angle oscillations were observed in multilayered conductors of organic origin (see, for example the articles \cite{9,10,11,12,13,14,15,16,17,18}) and in different quasi-two-dimensional conductors. This oscillation effect is essentially expressive under the $\tan\vartheta \gg 1$ condition, where the cross section of the FS with the plane $p_B=(\vec{p}\vec{B})/B=\textrm{const}$ is strongly extended along the axis $p_z=\vec{p}\vec{n}$, and the velocity of the electrons,~$v_z$, along the trajectories in the momentum space $\varepsilon(\vec{p})=\textrm{const}$, $p_B=\textrm{const}$, frequently changes the sign. Under certain orientations of the magnetic field, the mean value of the velocity $v_z$ may be small enough to cause a sharp increase of the current resistance which is repeated periodically as $\tan\vartheta$ functions. Out of the period of these oscillations, an important information on the form of FS could be gained \cite{19,20,21,22}. The inverse problem, i.e., calculation of the electron energy spectrum, is most effectively performed using experimental investigations of thermomagnetic effects in a strong magnetic field. An enormous number of the theoretical and experimental works are devoted to the investigation of the Nernst-Ettingshausen effect (see, for example \cite{23,24,25}).

\section{Formulation of the problem}
In the present report we consider the linear response of an electronic system to the perturbation by an electric field, $\vec{E}$, and temperature gradient,
$\partial T/\partial r$:
\begin{equation}\label{Eq.1}
J_i=\sigma_{ij}E_j-\alpha_{ij}\frac{\partial T}{\partial x_j}
\end{equation}
in layered conductors. The quasi-two-dimensional electronic energy spectrum of charge carriers is taken in a general form:
\begin{equation}\label{Eq.2}
\varepsilon(p)=\sum_{n=0}^\infty{\varepsilon_n{(p_x,p_y)}\cos\left[\frac{anp_z}{\hbar}+\alpha_n(p_x,p_y)\right]},
\end{equation}
\[\varepsilon_n{(-p_x,-p_y)}=\varepsilon_n{(p_x,p_y)}, \qquad \alpha_n{(-p_x,-p_y)}= \alpha_n{(p_x,p_y)}.\]
In the above equation, $a$ is the separation between the layers, $\hbar$ is Plank constant divided by $2\pi$, while $\varepsilon_n{(p_x,p_y)}$ and $\alpha_n{(p_x,p_y)}$ are arbitrary functions. The quasi-two-dimensionality parameter of the electronic energy spectrum, $\eta$, is defined as a ratio between the maximum value of the electron velocity along the normal of the layers:
\begin{equation}\label{Eq.3}
v_z=-\sum_{n=1}^\infty {\frac{an}{\hbar}\varepsilon_n{(p_x,p_y)}\sin\left[{\frac{anp_z}{\hbar}+\alpha_n{(p_x,p_y)}}\right]} \leqslant \eta v_\textrm{F}
\end{equation}
and the Fermi velocity $v_\textrm{F}$ with which the electrons move within the layer.

In the absence of the current density, the electric field generated by the temperature gradient, is given by
\begin{equation}\label{Eq.4}
E_i=\varrho_{ik}\alpha_{kj}\frac{\partial T}{\partial {x_j}}\,,
\end{equation}	  	
where $\varrho_{ik}$, $\sigma_{ij}$ and $\alpha_{ij}$ are the resistivity tensor, conductivity tensor and thermoelectricity tensor, respectively:
\begin{equation}\label{Eq.5}
\sigma_{ij}=-\int{\sigma_{ij}{(\varepsilon)}\frac{\partial f_\textrm{0}}{\partial\varepsilon} \rd\varepsilon},
\end{equation}
\begin{equation}\label{Eq.6}
\alpha_{ij}=-\int{\sigma_{ij}{(\varepsilon)}\frac{\varepsilon-\mu}{T}\frac{\partial f_\textrm{0} {(\varepsilon)}}{\partial\varepsilon} \rd\varepsilon}.
\end{equation}
Here,
\[f_\textrm{0}{(\varepsilon)}=\left[1+\exp\left(\frac{\varepsilon-\mu}{T}\right)\right]^{-1}\]
is the equilibrium Fermi distribution function for the charge carriers, $\mu$  is the chemical potential of the electrons, and $T$ is temperature in units of energy.

\begin{figure}[!b]
\centerline{\includegraphics[width=0.4\textwidth]{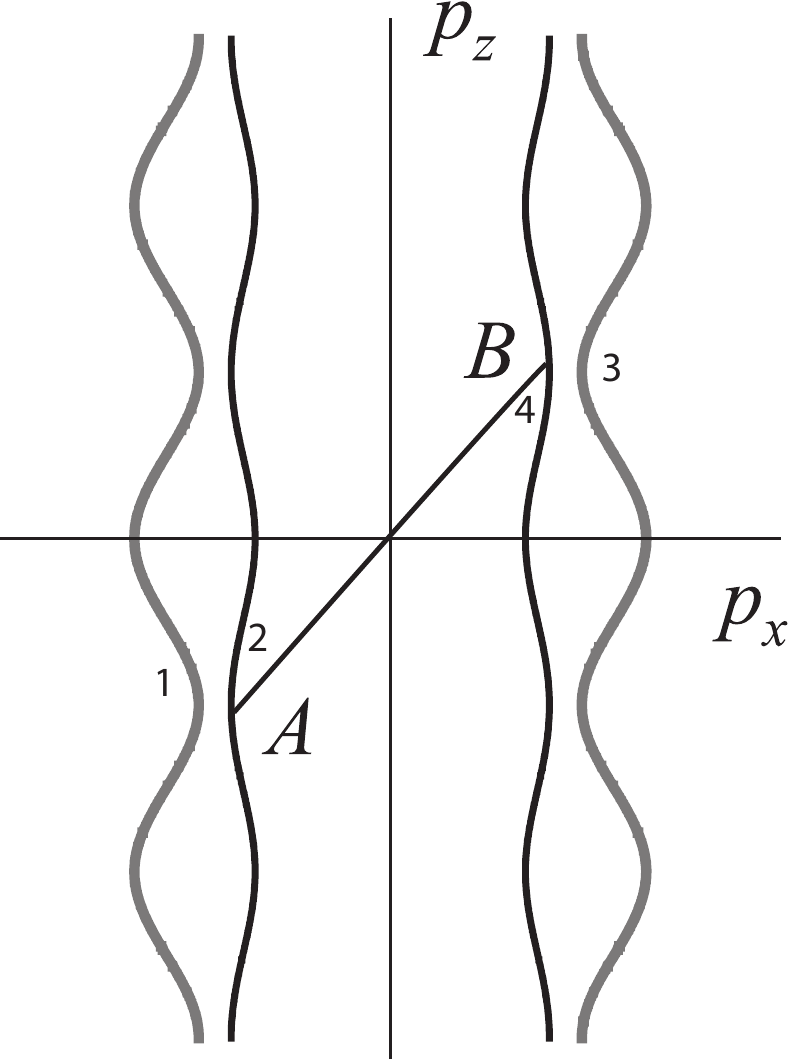}}
\caption{Projections of the FS and the magnetic breakdown electron trajectories in the magnetic field  $\vec{H}=(H\sin\vartheta,0,H\cos\vartheta)$
on the plane $p_{x}p_{y}$.} \label{fig1}
\end{figure}

By employing the kinetic equation solution for the distribution function of the charge carriers
$$f{(\vec{p})}=f_\textrm{0}{(\varepsilon)}-eE_j\psi^j\frac{\partial f_\textrm{0}{(\varepsilon)}}{\partial\varepsilon}$$
in the $\tau$-approximation for the collision integral, it is possible to calculate the  conductivity tensor
$\sigma_{ij}{(\varepsilon)}$:
\begin{equation}\label{Eq.7}
\sigma_{ij}{(\varepsilon)}=\frac{2e^3B}{c(2\pi\hbar)^3}\int{\rd p_B}\int_{0}^T{\rd t v_i(t)}\left[\,\,\int_{\lambda_\textrm{1}}^t{\rd t' v_j(t')+\exp\left(\frac{\lambda_\textrm{1}-t}{\tau}\right)\psi^j{(\lambda_\textrm{1},p_B)}}\right]=\langle{v_i \psi^j}\rangle,
 \end{equation}
where the function
\begin{equation}\label{Eq.8}
\psi^j{(\lambda_\textrm{1}, p_H)}=\int_{-\infty}^{\lambda_\textrm{1}}{v_j(t)\exp\left(\frac{t-\lambda_1}{\tau}\right)\rd t}
\end{equation}
describes the complex motion of the charge carriers along the magnetic breakdown trajectories, with probabilities for a magnetic breakdown $w$ and $w'$ in the regions A and B (figure~\ref{fig1}), respectively, at the moments  $\lambda_\textrm{1}$, $\lambda_\textrm{2}$, $\lambda_\textrm{3}$, ($\lambda_\textrm{1}$ being the closest to the moment when electrons move from a given sheet of FS to the neighboring one, and also $\lambda_k > \lambda_{k+1}$). The tensor $\sigma'_{ij}{(\varepsilon)}$ coincides with the tensor $\sigma_{ij}{(\varepsilon)}$ if  in the expression for $\sigma'_{ij}{(\varepsilon)}$, $\tau$ is substituted by $\tau_\varepsilon$.

As a result of an external generalized force, the FS sheets in the layered conductors with a multisheet FS appear to be close enough so that the charge carriers (as a result of magnetic breakdown) can move from one FS sheet to another, and their motion along the magnetic-breakdown trajectories becomes complex.
The magnetic breakdown oscillations of the current resistance normal to the layer of  conductors with a multisheet FS, composed of a weakly corrugated cylinder and two corrugated planes, is periodically repeated in the momentum space, were theoretically investigated in the work \cite{26}.
It was shown that the period of the magnetic breakdown oscillations of the kinetic coefficients contains important information for both the form of the FS planes (sheets) and the mutual positions of the FS planes (sheets) in the momentum space. Furthermore, only the limiting cases when the probability for magnetic breakdown,~$w$,  is negligibly small and when $w$ is close to 1, are considered.

In this report we consider the thermoelectric effects under a general assumption for the value of the magnetic breakdown probability. Let the Fermi surface be composed of a cylinder and two planes, weakly corrugated along the $p_z$ axis, and let the $p_x$ axis be normal to the plane, figure \ref{fig1}.

When there are several groups of charge carriers, each of them contributes to the kinetic coefficient, so that
\begin{equation}\label{Eq.9}
\langle{v_i\psi^j}\rangle=\langle{v_i\psi^j}\rangle_\textrm{1}+\langle{v_i\psi^j}\rangle_\textrm{2}+\langle{v_i\psi^j}\rangle_3+\langle{v_i\psi^j}\rangle_4,
\end{equation}
where the terms ${\langle{v_i\psi^j}\rangle}_\textrm{1}+{\langle{v_i\psi^j}\rangle}_\textrm{3}$   define the current due to the electrons on the quasi-planar FS sheets and the terms ${\langle{v_i\psi^j}\rangle}_\textrm{2}+{\langle{v_i\psi^j}\rangle}_\textrm{4}$ give the contribution of the conduction electrons belonging to the FS section in the form of a corrugated cylinder.

\section{Calculations}
In \cite{27}, a case was considered where the probabilities for magnetic breakdown $w$ and $w'$ are essentially different, and the electron that was tunneled from the quasi FS plane sheet to the cylinder FS sheet, after performing a rotation along the cross-section (as a result of magnetic breakdown) comes back to the previous quasi FS plane sheet.

Here, we consider the case where in the magnetic field $\vec{B} = {(B\sin\vartheta,0,B\cos\vartheta)}$ the distance between the regions A and B (when the  FS sheets are at the closest distance along the $p_z$ axis) is equal to an integer,~$N$, of the unit cells in the momentum space, figure \ref{fig1}, i.e.,
\begin{equation}\label{Eq.10}
\tan\vartheta=\frac{2\pi N\hbar}{aD_{p}}\,.
\end{equation}
In the above equation, $D_{p}$ is the diameter of a cylinder along the $p_x$ axis. In this case, the probability~$w'$  coincides with the probability $w$ on all cross-sections of the FS and the plane $p_B=\textrm{const}$, and the electron could move from one quasi-FS plane sheet to another quasi-FS plane sheet even in cases where the corrugation along the $p_x$ axis is pronounced.

Non-equilibrium distribution functions of the electrons on the FS sheet 1 after the magnetic breakdown in the region A:
\begin{equation}\label{Eq.11}
\phi_\textrm{1}{(\lambda_\textrm{1}+0)}=\int_{-\infty}^{\lambda_\textrm{1}}\rd t\exp{\left(\frac{t-\lambda_\textrm{1}}{\tau}\right)}\frac{{(\vec{E}\vec{v}})_\textrm{1}}{E}
\end{equation}
are connected to the distribution functions of the electrons on the same FS sheet but prior to the magnetic breakdown, $\phi_1{(\lambda_1-0)}$, with the following equation:
\begin{equation}\label{Eq.12}
\phi_\textrm{1}{(\lambda_\textrm{1}+0)}=(1-w) \phi_\textrm{1}{(\lambda_\textrm{1}-0)}+ w\phi_\textrm{2}(\lambda_\textrm{1}-0),
\end{equation}
and with the functions $\phi_\textrm{1}$ and $\phi_\textrm{2}$ at the previous moment $\lambda_\textrm{2}$
\begin{equation}\label{Eq.13}
\phi_\textrm{1}{(\lambda_\textrm{1}+0)}=(1-w)\left[A_\textrm{1}+\exp{\left(\frac{-P}{\tau}\right)}\phi_\textrm{1}{(\lambda_\textrm{2}+0)}\right]+w\left[A_\textrm{2}+\exp{\left(\frac{-P\,'}{\tau}\right)}\phi_\textrm{2}{(\lambda_\textrm{2}+0)}\right] .
\end{equation}
Similarly, the equation (\ref{Eq.13}) at the earlier moments $\lambda_\textrm{2}$, $\lambda_\textrm{3}$, $\lambda_\textrm{4}$, gain the forms:
\begin{equation}\label{Eq.14}
\phi_\textrm{2}{(\lambda_\textrm{2}+0)}=(1-w)\left[A_\textrm{4}+\exp{\left(\frac{-P\,'}{\tau}\right)}\phi_\textrm{4}{(\lambda_\textrm{3}+0)}\right]+w\left[A_\textrm{3}+\exp{\left(\frac{-P}{\tau}\right)}\phi_\textrm{3}{(\lambda_\textrm{3}+0)}\right],
\end{equation}
\begin{equation}\label{Eq.15}
\phi_\textrm{3}(\lambda_\textrm{3}+0)=(1-w)\left[A_\textrm{3}+\exp{\left(\frac{-P}{\tau}\right)}\phi_\textrm{3}(\lambda_\textrm{4}+0)\right]+w\left[A_\textrm{4}+\exp{\left(\frac{-P\,'}{\tau}\right)}\phi_\textrm{4}(\lambda_\textrm{4}+0)\right],
\end{equation}
\begin{equation}\label{Eq.16}
\phi_\textrm{4}(\lambda_\textrm{4}+0)=(1-w)\left[A_\textrm{2}+\exp{\left(\frac{-P\,'}{\tau}\right)}\phi_\textrm{2}(\lambda_\textrm{5}+0)\right]+w\left[A_\textrm{1}+\exp{\left(\frac{-P}{\tau}\right)}\phi_\textrm{1}(\lambda_5+0)\right],
\end{equation}
\begin{equation}\label{Eq.17}
\phi_\textrm{1}(\lambda_\textrm{5}+0)=(1-w)\left[A_\textrm{1}+\exp{\left(\frac{-P}{\tau}\right)}\phi_\textrm{1}(\lambda_\textrm{6}+0)\right]+w\left[A_\textrm{2}+\exp{\left(\frac{-P\,'}{\tau}\right)}\phi_\textrm{2}(\lambda_6+0)\right].
\end{equation}
In the above equations, the functions
\begin{equation}\label{Eq.18}
A_i=\int_{\lambda_j+1}^{\lambda_j}\rd t\exp{\left(\frac{t-\lambda_j}{\tau}\right)}\frac{{(\vec{E}\vec{v}})_i}{E}\,, \qquad  i=1,2,3,4,
\end{equation}
in the non-collision range ($\tau \to \infty$) are equal to the drifting of the electrons along an electrical field for a time period (${\lambda_j-\lambda_{j+1}}$) between two instants of the magnetic breakdown. This period  estimated with accuracy to small corrections proportional to the quasi-two-dimensionality parameter of the electronic energy spectrum, $\eta$, is independent of $\lambda_j$, and corresponds to the period $P$ of the electron drifting on the FS sheets 1 and 3, i.e., to a half-period $P\,'$ of the electron drifting along the closed sections of the corrugated cylinder.

One can easily find that the equation (\ref{Eq.17}) corresponds to the equation (\ref{Eq.13}) for some previous magnetic breakdown moment. Going on with the above recursive relation, one moves back in time and, since the functions on the right-hand side of the equations  (\ref{Eq.12})--(\ref{Eq.17}) decrease with each recursion (accumulation of the effect of multipliers smaller than 1), they become sufficiently small after many recursive steps. As a result, the functions $\phi_i$  on the left-hand side of the equations (\ref{Eq.13})--(\ref{Eq.16}), proportional to $A_j$, form a geometric progression which can be easily calculated \cite{24}. After several algebraic manipulations, one comes to
\begin{equation}\label{Eq.19}
\phi_\textrm{1}{(\lambda_\textrm{1}+0)}=\frac{(1-w)A_\textrm{1}+wA_\textrm{2}}{1-h_\textrm{1}}+\sum_{n=0}^\infty{h_\textrm{1}^n g\phi_\textrm{2}(\lambda_{n+2}+0)},
\end{equation}
where  $h_\textrm{1}=(1-w)\exp({-P/\tau})$, $ g=w\exp{(-P\,'/\tau)}$.

Substituting  the function $\phi_4{(\lambda_4+0)}$ in the equation (\ref{Eq.14}) and using the expression (\ref{Eq.16}), one gets
\begin{equation}\label{Eq.20}
\phi_\textrm{2}(\lambda_\textrm{2}+0)=(1-w)(A_\textrm{4}+hA_\textrm{2})+w(A_\textrm{3}+hA_\textrm{1})+h^2\phi_\textrm{2}(\lambda_4+0)
+g_\textrm{1}\phi_\textrm{3}(\lambda_\textrm{3}+0)+hg_\textrm{1}\phi_\textrm{1}(\lambda_\textrm{5}+0),
\end{equation}
where $g_\textrm{1}=w\exp{(-P/\tau)}$, $ h=(1-w)\exp{(-P\,'/\tau)}$.

Using the expression (\ref{Eq.16}), the connection between the function $\phi_\textrm{3}(\lambda_3+0)$:
\begin{equation}\label{Eq.21}
\phi_\textrm{3}(\lambda_3+0)=\frac{(1-w)A_\textrm{3}+wA_\textrm{4}}{1-h_\textrm{1}}+\sum_{n=0}^\infty{h_\textrm{1}^n g\phi_\textrm{4}(\lambda_{n+4}+0)} ,
\end{equation}
and the function $\phi_\textrm{2}$ is as follows:
\begin{equation}\label{Eq.22}
\phi_\textrm{3}(\lambda_3+0)=\frac{(1-w)A_\textrm{3}+wA_\textrm{4}}{1-h_\textrm{1}}+hg\frac{(1-w)A_\textrm{2}+wA_\textrm{1}}{1-h_\textrm{1}}+
\sum_{n=0}^\infty{h_\textrm{1}^n  gh\phi_\textrm{2}(\lambda_{n+5}+0)}+ \sum_{n=0}^\infty{h_\textrm{1}^n  gg_\textrm{1}\phi_\textrm{1}(\lambda_{n+5}+0)}.
\end{equation}

By substituting the functions $\phi_\textrm{1}$ and $\phi_\textrm{3}$ from the equations (\ref{Eq.21})--(\ref{Eq.22}) in the equation (\ref{Eq.20}), we gain the functional expression for only one function $\phi_\textrm{2}$  whose solution under conditions
$w \gg \gamma_\textrm{1}=\exp{(P/\tau)}-1$  and $w \gg \gamma=\exp{(P\,'/\tau)}-1$  gains a simple form:
\begin{equation}\label{Eq.23}
\phi_\textrm{2}(\lambda_\textrm{2}+0)=\frac{A_\textrm{1}+A_\textrm{2}+A_\textrm{3}+A_\textrm{4}}{2(\gamma+\gamma_\textrm{1})}\,,
\end{equation}
\begin{equation}\label{Eq.24}
\phi_\textrm{1}(\lambda_\textrm{1}+0)=\frac{(1-w)A_\textrm{1}+wA_\textrm{2}}{w+\gamma_\textrm{1}}+\frac{w}{w+\gamma_\textrm{1}}\phi_\textrm{2}(\lambda_\textrm{2}+0).
\end{equation}
The functions $\phi_\textrm{3}(\lambda_\textrm{3}+0)$ and $\phi_\textrm{4}(\lambda_\textrm{4}+0)$ coincide with the functions $\phi_\textrm{1}(\lambda_\textrm{1}+0)$ and $\phi_\textrm{2}(\lambda_\textrm{2}+0)$ if we interchange the places therein: $A_\textrm{1} \to A_\textrm{3}$,  $A_\textrm{2} \to A_\textrm{4}$. Now, if we know the functions $\phi_i$, it is easy to calculate all the components of the conductivity and thermoelectricity tensors under different values and orientations of the magnetic field $\vec{B}=(B\cos\varphi\sin\vartheta, B\sin\varphi\sin\vartheta, B\cos\vartheta)$.

\section{Discussion and conclusion}
The basic contribution to the average value of  an electron velocity, $v_z$, that is moving along a strongly extended trajectories, under condition $\tan\vartheta \gg 1$, is given by small areas in the vicinity of the stationary phases, where
\begin{equation}\label{Eq.25}
\frac{\rd p_z}{\rd t}=\frac{eB}{c}\sin{\vartheta}{(v_x\sin{\varphi}-v_y\cos{\varphi})}=0.
\end{equation}
On the closed sections of a corrugated cylinder, there are at least two points of this kind. Under certain orientations of the magnetic field with respect to the crystal axis of the single crystal conductor, the contribution from these points to the average value of the electron velocity, $v_z$, can be cancelled. This leads to a sharp increase of the current resistance normal to the layer, $\varrho_{zz}$, whose asymptotic behavior under $\eta \to 0$  is equal to $ 1/\sigma_{zz}$. In the magnetic field normal to the $y$-axis, the components of the conductivity tensor gain the form:
\begin{eqnarray}\label{Eq.26}
&&\sigma_{zz}=\frac{\sigma_\textrm{0}\eta^2}{\tan\vartheta}\left\{\beta (1+\sin{\alpha} D_{p})+2\beta_\textrm{1} (1+\sin{\alpha}\delta p_x)+\beta_\textrm{2} \left[2\cos\alpha{(D_{p}+\delta p_x+\Delta_{p})}\right.\right.
\nonumber \\&& \qquad \quad \qquad\quad+\left.\sin\alpha{(D_{p}+2\delta p_x+2\Delta_{p})}-\sin\alpha(D_{p}+\delta p_x)\right]+\beta_\textrm{3} \left[\cos\alpha{(\delta p_x+\Delta_{p})}\right.
\nonumber \\&&\qquad \qquad \quad\quad\left.\left.-\sin\alpha{\Delta_{p}}+\sin\alpha{(D_{p}+2\delta p_x+\Delta_{p})}+\cos\alpha{(D_{p}+\Delta_{p})}\right]\right\}.
\end{eqnarray}

Here, $\sigma_\textrm{0}$ is the electro-conductivity of the quasi-two-dimensional conductor along the layer in the absence of a magnetic field; $D_{p}$  is the diameter of a cylinder along  the $p_x$  axis; $\Delta_{p}=p_{x2}^\textrm{min}-p_{x1}^\textrm{min}= p_{x3}^\textrm{min}-p_{x2}^\textrm{min}$ is the minimum distance between the cylinder and the plane FS sheets;  $\alpha=(a/\hbar)\tan\vartheta$; the quantities $\beta$, $\beta_\textrm{1}$, $\beta_\textrm{2}$ and $\beta_\textrm{3}$ are all $\sim 1$ and depend on a concrete type of the electronic energy spectrum. In the formula~(\ref{Eq.26}), the non-oscillation terms of the diagonal electro-conductivity components $\sigma_{zz}$  are not included:
\begin{equation}\label{Eq.27}
\sigma_{zz}=-\frac{2e^2}{(2\pi\hbar)^3}\int{\rd\varepsilon\frac{\partial f_\textrm{0} {(\varepsilon)}}{\partial\varepsilon}}\int{\rd p_B\left\vert{\frac{tB}{c}}\right\vert\frac{(\bar{v}_{z1}+\bar{v}_{z2}+\bar{v}_{z3}+\bar{v}_{z4})^2}{2(\gamma+\gamma_1)}}>0.
\end{equation}                              	
The angular oscillation of a thermoelectric filed along the normal to the layer, $E_z$:
\begin{equation}\label{Eq.28}
E_z=\frac{\pi^2 T}{3e}\varrho_{zk}\frac{\partial\sigma_{kj}}{\partial\mu}\frac{\partial T}{\partial x_j}
\end{equation}
gains a multiplier proportional to $\tan\vartheta$ as a result of the differention with respect to $\mu$ of the quickly oscillating members in $\sigma'_{ij}(\varepsilon)$ under condition $\tan\vartheta \gg 1$. It means that the thermoelectric filed, $E_z$, as a function of $\tan\vartheta$, can change the sign even in the case of a longitudinal thermoelectrical effect when the temperature gradient is directed along the normal of the layer. When the magnetic field is deflected from the $xz$-plane by the angle $\varphi$ different from zero, the drift of the charge carriers along the $y$-axis is $\bar{v}_y=\bar{v}_z\sin\varphi\tan\vartheta$,
and, when $\sin\varphi\tan\vartheta \gg 1$, the oscillation part of the thermoelectrical field $E_z$:
\begin{equation}\label{Eq.29}
E_z=\frac{\pi^2 T}{3e}\frac{\tau}{\tau_{\eta}}\frac{1}{\sigma_{zz}}\frac{\partial\sigma_{zz}^\textrm{osc}}{\partial\mu}\left({\frac{\partial T}{\partial z}+\sin\varphi\tan\vartheta\frac{\partial T}{\partial y}}\right)
\end{equation}
is mainly determined with the Nernst-Ettingshausen effect.
The magnetic breakdown oscillations could not exist if $\sin\varphi \cong 1$ because in this case the relation (\ref{Eq.25}), which is an imperative requirement for the existence of stationary phase points on the FS plane sheets, is not fulfilled. Under the following conditions: $\gamma_\textrm{0} \ll \cos\vartheta \ll \sin\varphi \ll 1$, (where
$\gamma_\textrm{0}$ is of the same order of magnitude with the quantities $\gamma_\textrm{1}$ and $\gamma_\textrm{2}$ under
$\vartheta=0$), the stationary phase points on the electronic trajectories, under small  deflection of the
magnetic field from the $xz$-plane,  are insignificantly dislocated. This allows us to use the equation
(\ref{Eq.26}) to calculate the angular oscillation of a thermoelectrical filed along the normal to the layer,
$E_z$.
There are also angular oscillations of the thermoelectrical filed along the layer plane under
condition $\tan\vartheta \gg 1$, but their  amplitudes are negligibly small (proportional to $\eta^2$) compared to the dominant background, smoothly changing with the variation of the angle between the magnetic field and the layer plane.

\ukrainianpart

\title[]
{Осциляції магнітного пробою поля Нерста-Еттінгсгаузена в  шаруватих провідниках}

\author[]{O.~Галбова}

\address{Факультет природничих наук і математики, Інститут фізики, \\P.O. Box 162 1001 Скоп'є, Республіка Македонія}

\makeukrtitle

\begin{abstract}

В цій статті досліджується ефект Нерста-Еттінгсгаузена в шаруватих провідниках. Розглядаючи поверхню Фермі  (ПФ), що складається зі слабо   гофрованого циліндра і двох гофрованих площин, розподілених періодично в імпульсному просторі, вивчаються термоелектричні ефекти, роблячи загальні припущення для значень ймовірності магнітного пробою. В результатi зовнiшньої узагальненої сили, шари ПФ в шаруватих провiдниках з
 багатошаровими ПФ виникають так близько один до одного, що носiї заряду (в результатi магнiтного пробою) можуть переходити з одного шару ПФ до iншого. Крім того, обчислено функції розподілу носіїв заряду та осциляції термоелектричного поля магнітного пробою вздовж нормалі до шару для різних значень і орієнтацій магнітного поля  $B$. Показано, що якщо магнітне поле відхиляється від  $xz$-площини на кут $\varphi$, осциляційна частина термоелектричного поля вздовж нормалі до шару при умові $\sin\varphi\tan\vartheta \gg 1$ визначається в основному ефектом Нерста-Еттінгсгаузена.

\keywords шаруватий провідник, поверхня Фермі, осциляції магнітного пробою

\end{abstract}

\end{document}